\documentclass[twocolumn,showpacs,prb,amsmath,amssymb,floatfix]{revtex4}

\usepackage{graphicx}
\usepackage{dcolumn}
\usepackage{bm}
\usepackage{latexsym}
\usepackage[usenames]{color}
\usepackage{amsmath}

\makeatletter
\def\lsim{\mathrel{\mathpalette\gls@align<}}
\def\gsim{\mathrel{\mathpalette\gls@align>}}
\def\gls@align#1#2{\lower.6ex\vbox
	{\baselineskip\z@skip\lineskip\z@\ialign
		{$\m@th#1\hfill##$\crcr#2\crcr\sim\crcr}}}
\makeatother
\def\be{\begin{equation}}
\def\ba{\begin{align}}

\begin{document}
\preprint{Complete homochirality induced by the 
nonlinear 
autocatalysis and recycling
}
\title{Complete homochirality induced by the 
nonlinear 
autocatalysis and recycling
}

\author{Yukio Saito}
 \email{yukio@rk.phys.keio.ac.jp}
\author{Hiroyukio Hyuga}
 \email{hyuga@rk.phys.keio.ac.jp}
\affiliation{
	Department of Physics, Keio University,
		3-14-1 Hiyoshi, Kohoku-ku, Yokohama 223-8522, Japan
}


\begin{abstract}
A nonlinear autocatalysis of a chiral substance is shown 
to achieve homochirality in a closed system, if the  back-reaction
is included. 
Asymmetry in the concentration of 
two enantiomers or the enantiometric excess increases due to the
nonlinear autocatalysis. 
Furthermore, 
when the back-reaction is taken into account, the reactant
supplied by the decomposition of the enantiomers is recycled to produce
more and more the dominant one,
and eventually the homochirality is established. 
\end{abstract}

\maketitle

Natural organic molecules 
associated with living matters usually have two possibilities in their stereo-structures, in a right-handed form(R) or in a mirror-image left-handed form(S).
\cite{bonner88,feringa+99}
These two forms are called enantiomers, and the molecules are said to be 
chiral.
Chiral molecules are optically active to rotate the direction of polarization of plane polarized light.
From the energetic point of view, these two enantiomers can exist 
with an equal probability,
but 
the life on earth utilizes only one type:
 only levorotatory(L)-amino acids  (or S)
and 
dextrorotatory(D)-sugards (or R).
This symmetry breaking in the chirality is called the homochirality.

The origin of this unique chirality has long intrigued many scientists.
\cite{bonner88}
In order to find the physical origin of this homochirality,
initial asymmetry in the primordial molecular environment has to be created
by chance or engendered deteminately by external or internal factors,
such as the parity breaking effect in the weak interaction,
\cite{mason+85,kondepudi+85,meiring87,bada95}
the asymmetry in circularly polarized light,\cite{bailey+98,feringa+99}
or adsorption on optically active crystals.\cite{hazen+01}
Then the induced small initial chiral asymmetry has to be amplified. 

Frank has shown theoretically that an autocatalytic reaction 
of a chemical substance with an antagonistic process
can lead to an amplification of enantiometric excess (ee)
and  to homochirality.\cite{frank53}
Many theoretical models are proposed afterwards, 
but they are often criticized as lacking any experimental support.
\cite{bonner88}
Recently, asymmetric autocatalysis
\cite{wynberg89}
 of pyrimidyl alkanol 
 has been
studied intensively.
\cite{soai+90,soai+95,sato+01,sato+03}
The enhancement of ee 
was
 confirmed,\cite{soai+95} and
its temporal evolution 
was
explained by the second-order autocatalytic
reaction.\cite{sato+01,sato+03} 
But only with the nonlinear autocatalysis,
chirality selection is not complete
and the value of ee stays less than unity.
Here we show that the complete homochirality is achieved by the
back-reaction to recycle the reactant.
If the rate of back reaction is small, it takes a long time before
the homochirality is achieved.


We consider a chemical reaction such that substances A and B 
react to form substance C. Though reactants A and B are achiral,
the product C happened to be chiral in two enantiometric forms; 
$R$-isomer ($R$)-C and $S$-isomer ($S$)-C.
\begin{align}
{\mathrm A} +{\mathrm B} \rightleftharpoons (R){\mathrm -C}
\nonumber \\
{\mathrm A} +{\mathrm B} \rightleftharpoons (S){\mathrm -C}
\end{align}

In the experiment by Soai et al.\cite{soai+95} 
one provides a small amount of 
enantiomers with concentrations $R_0$ and $S_0$ respectively in the
reactants of concentrations $A_0$ and $B_0$.
Instead of the open flow system considered by Frank,\cite{frank53}
we assume a closed conserve system such that
concentrations $R$ and $S$ at time $t$ 
 vary in proportion to the present amount of the reactants A and B as
\begin{align}
\frac{dR}{dt} = k_R( A_0+R_0+S_0-R-S)(B_0+R_0+S_0-R-S)
\nonumber \\
\frac{dS}{dt} = k_S( A_0+R_0+S_0-R-S)(B_0+R_0+S_0-R-S).
\end{align}
Here we first neglect the back reaction from $(S)$-C or $(R)$-C to A and B.
Later, the effect of back reaction is shown to be very important in 
achieving the complete homochirality.

The positive reaction coefficients, $k_{R.S}$, may depend on the concentrations 
$R$ and $S$, if the reaction is autocatalytic.
We consider various cases step by step in the following.
Furthermore, for a simplicity reason we assume
that there 
is an
 ample amount of substance B, 
and the reaction is controlled by the less abundant substance A.
Then by normalizing the concentration by the initial concentration 
$A_0+R_0+S_0$ as $r=R/(A_0+R_0+S_0)$, $s=S/(A_0+R_0+S_0)$,  
the equation is  simplified as
\begin{align}
\frac{dr}{dt} = k_r( 1-r-s)
\nonumber \\
\frac{ds}{dt} = k_s( 1-r-s)
\end{align}
where the reaction rate coefficient 
$k_{r,s}=k_{R,S}(B_0+R_0+S_0)$ may depend on $r$ and $s$.

Since both enantiomers are equivalent except the handedness, the time evolution
should be symmetric in $r$ and $s$. The chirality of the system can be brought
about through the asymmetry of the initial concentrations $r_0$ and $s_0$
of two enantiomers, ($R$)-C and ($S$)-C.
In the following we study the 
temporal variation of the enatiometric excess measured by
\begin{align}
\Omega = \frac{r-s}{r+s} .
\end{align}

{\it Non-Autocatalytic Case}

If the reaction is not autocatalytic, the reaction coefficients are simply
constant as $k_r=k_s=k$.
Then the ratio of the two equations (3) gives
$ {dr}/{ds} = 1 $,
which defines the trajectory in the $r-s$ space.
It is easily solved as $r-s=r_0-s_0=$const. 
The time evolution of enantiomers is depicted in the flow diagram in Fig.1(a).
The flow is on the line with a unit slope, and terminates on a point of
crossing
 the diagonal line $r+s=1$; the reaction stops because
the  reactant A is consumed up in a closed system.
The diagonal line $r+s=1$ is in fact a set of stable fixed points.

\begin{figure}[h]
\begin{center} 
\includegraphics[width=0.43\linewidth, height=0.41\linewidth]{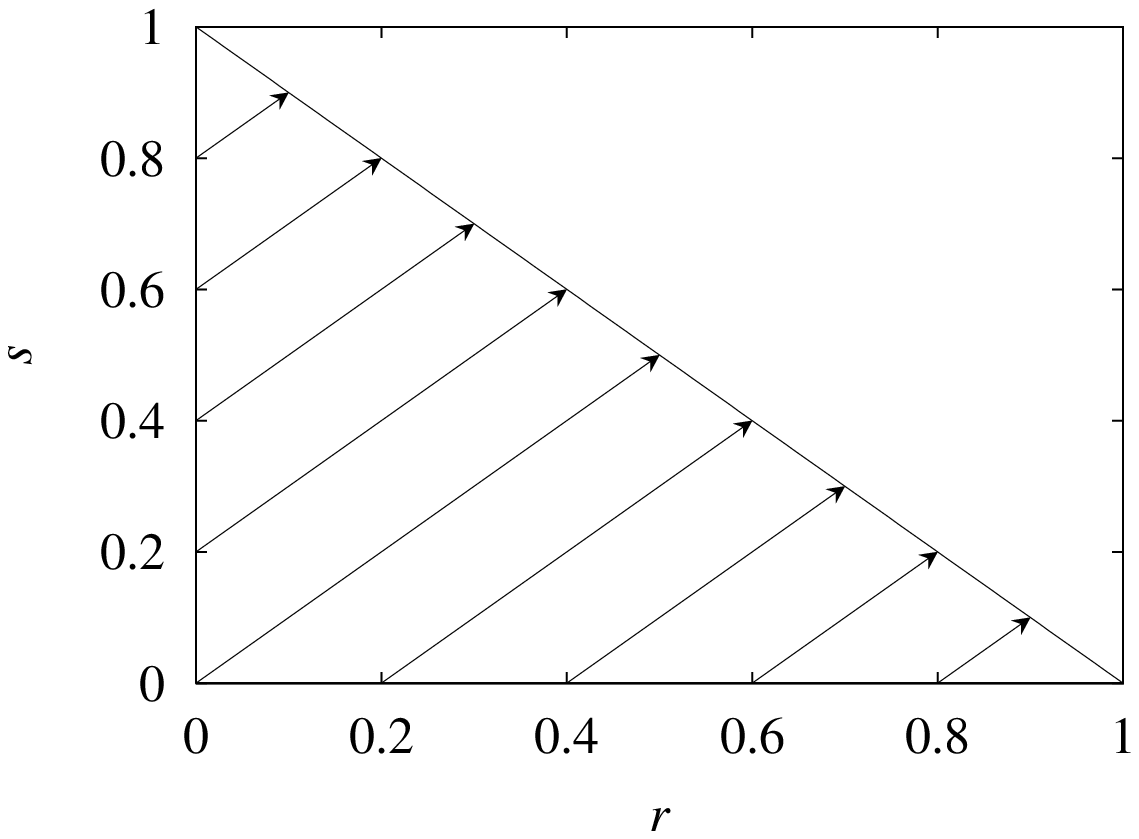}
\hspace{0.5cm}
\includegraphics[width=0.43\linewidth, height=0.41\linewidth]{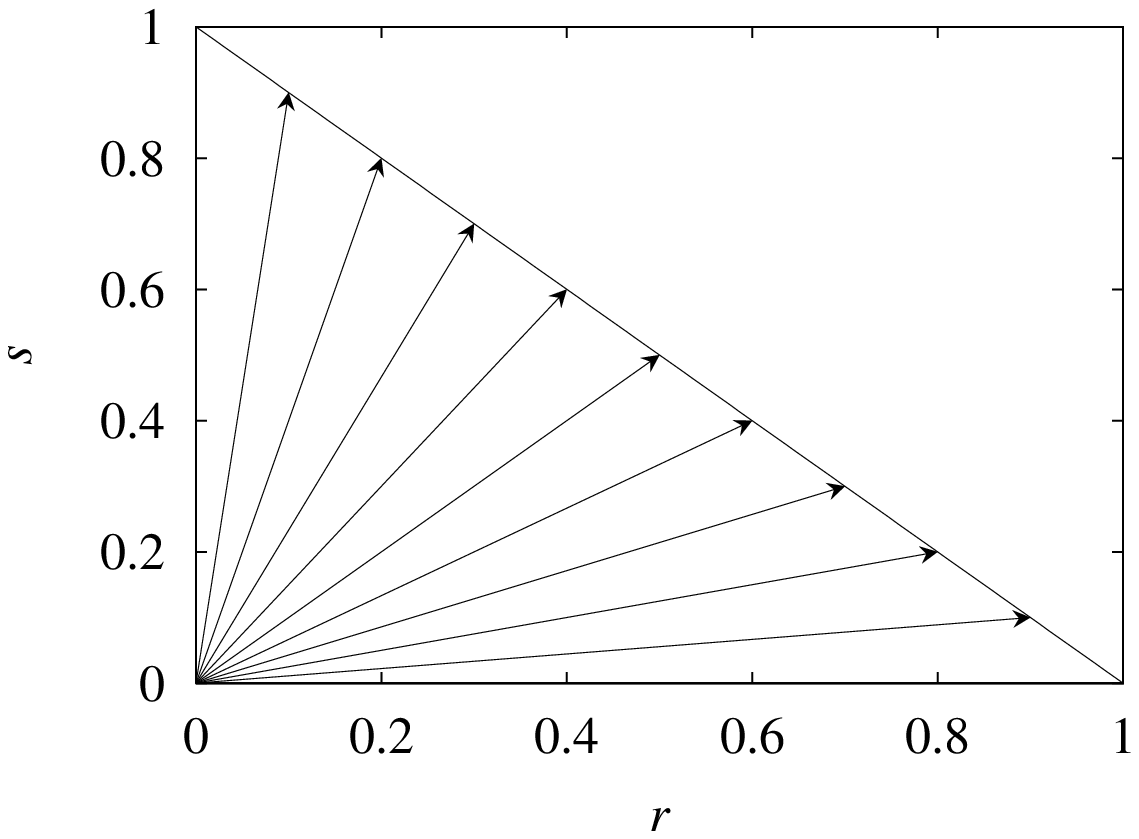}
\\
(a) Nonautocatalytic  
(b) Linearly autocatalytic 
\\
\includegraphics[width=0.43\linewidth, height=0.41\linewidth]{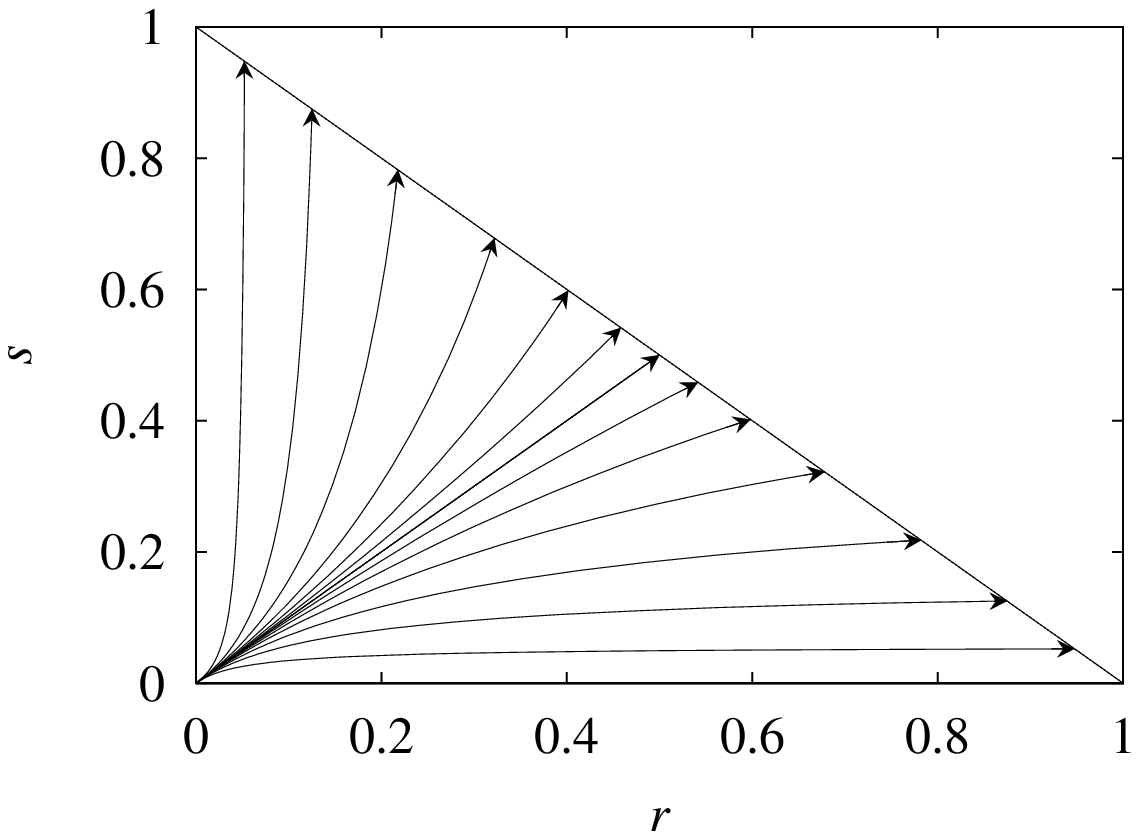}
\hspace{0.5cm}
\includegraphics[width=0.43\linewidth, height=0.41\linewidth]{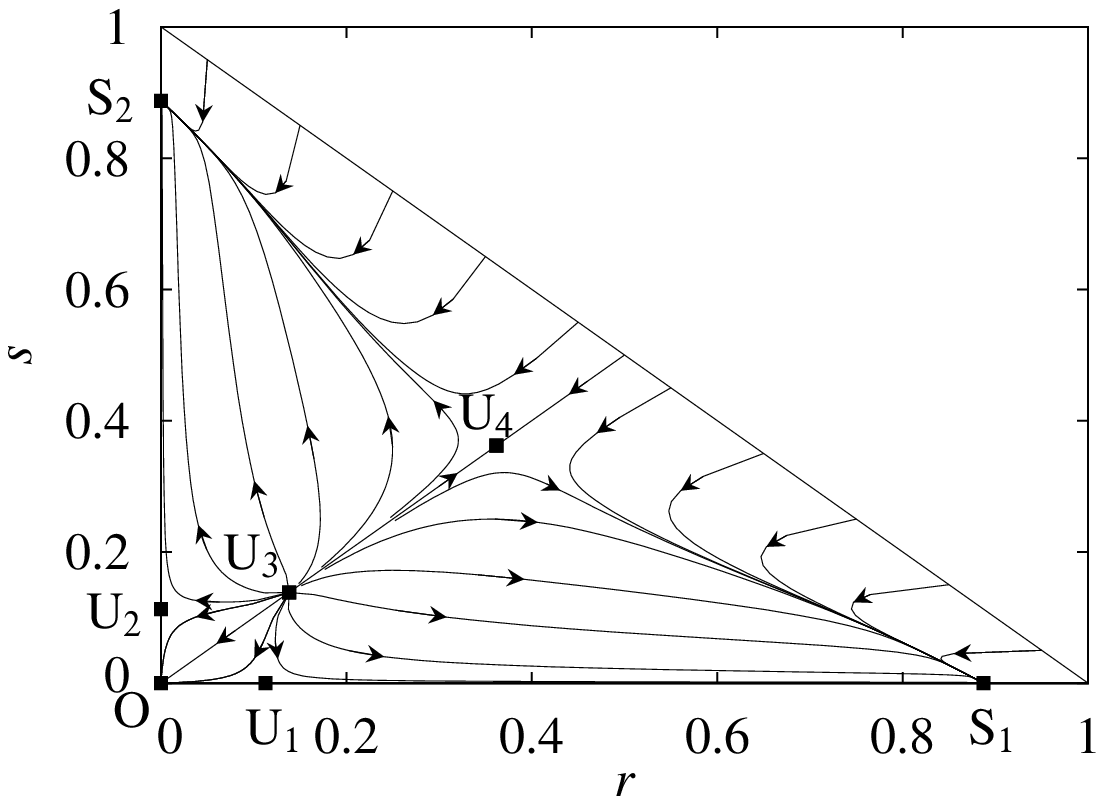}
\\
(c) Nonlinearly autocatalytic  
(d) 
with back reaction 
\end{center} 
\caption{Flow diagrams in $r$-$s$ space. (a) Nonautocatalytic, (b) linearly 
autocatalytic, (c) nonlinearly autocatalytic and (d) nonlinearly autocatalytic with backward reactions ($\lambda/k=0.1)$.
}
\label{fig1}
\end{figure}

The ee decreases since the numerator $r-s$ in $\Omega$
remains constant whereas the 
denominator $r+s$  increases in time.
Actually, the equation is easily solved, and ee is obtained as
$ \Omega = ({r_0-s_0})/({1-(1-r_0-s_0)e^{-2kt}}) .$
The ee decreases from the initial value $\Omega_0= (r_0-s_0)/(r_0+s_0)$ to
the asymptotic value $\Omega_{\infty} = r_0-s_0$, which  should be very small 
if the initial amounts 
$r_0=R_0/(A_0+R_0+S_0)$ and $s_0=S_0/(A_0+R_0+S_0)$ are small.
This corresponds to racemic mixtures.

{\it Linearly Autocatalytic Case}

If the reaction is autocatalytic in the first order, the reaction coefficient 
is linearly proportional to the product concentrations 
as $k_r=kr$ and $k_s=ks$.
In this case, the ratio of two rate equations gives
$ dr/ds={r}/{s} ,$
which is easily solved as
$r/s=r_0/s_0=$const. 
The time evolution is depicted in the flow diagram Fig.1(b).
The flow is radial from the origin  and trivially the
ee does not change even as the reaction proceeds,
\begin{align}
\Omega = \frac{r-s}{r+s} = \frac{r_0-s_0}{r_0+s_0} = \Omega_0
\end{align}
as first notified by Frank.\cite{frank53}

{\it Nonlinearly Autocatalytic Case}

If the reaction is autocatalytic in the higher order, the reaction coefficient 
depends nonlinearly on the product; for the second order autocatalysis, 
for example,
$k_r=kr^2$ and $k_s=ks^2$.
In this case, the ratio of two rate equations gives
\begin{align}
\frac{dr}{ds}=\frac{r^2}{s^2}
\end{align}
which is easily solved to give the trajectory in the $r-s$ space as
$1/r-1/s =1/r_0- 1/s_0=$const. 
The time evolution is depicted in the flow diagram Fig.1(c).
Below the diagonal $r=s$, the flow bends down, whereas above
the diagonal the flow bends up.
The final concentrations $r_{\infty}$ and $s_{\infty}$ 
should satisfy the relation $r_{\infty}+s_{\infty}=1$.
Then the final value of the ee is determined by the initial value $\Omega_0$
and the amount of the input $r_0$ (or $s_0$) as
\begin{align}
\Omega_{\infty} = r_{\infty}-s_{\infty} 
= {\mathrm{sgn}} (\Omega_0) \left( \sqrt{1+r_0^2 
\left(\frac{1-\Omega_0}{\Omega_0} \right) ^2}
- r_0 \frac{1-\Omega_0}{|\Omega_0|} \right)
\end{align}
If one starts with very small values of $r_0$ and $s_0$,
the ee approaches $|\Omega_{\infty}| \approx 1-r_0(1-\Omega_0)/|\Omega_0|$.
The amplification of ee is expected to be large 
 if the initial composition $r_0$ and $s_0$ are small.
For example, with the same initial ee as $\Omega_0=2\%$, if one starts
with  $r_0=0.051$ and $s_0=0.049$,  
the final ee is $\Omega_{\infty}=19\%$ according to eq.(6), whereas
with the initial values 
$r_0=0.00051$ and $s_0=0.00049$, then 
$\Omega_{\infty}=98\%$.

{\it Homochirality by Back Reaction}

From the previous analysis we found that the ee amplification takes place 
with a nonlinear or higher-order autocatalytic chemical reaction.
However, the final chirality is not complete. 
The ee is smaller than 100\%, even
the starting values $r_0$ and $s_0$ are very small.
Here we show that the inclusion of the back reaction from the 
products ($R$)-C and ($S$)-C to A 
brings about the drastic change in the flow diagram.

A simple back reaction process 
is incorporated into the rate equation (3) as
\begin{align}
\frac{dr}{dt}=kr^2(1-r-s)-\lambda r
\nonumber \\
\frac{ds}{dt}=ks^2(1-r-s)-\lambda s
\end{align}
with a rate constant $\lambda$ of the back reaction.
Now the flow diagram in $r-s$ space changes from Fig.1(c) to (d).
The diagonal line $r+s=1$ is no more a set of infinitely 
many stable fixed points.
Instead, there are seven fixed points in the flow diagram, Fig.1(d).
The trivial fixed point at the origin O:$(r_o,s_o)=(0,0)$ is stable.
On the $r$-axis there are two fixed points; 
U$_1$:$(1-\sqrt{1-4\lambda/k})/2,0)$
is unstable and S$_1$:$(1+\sqrt{1-4\lambda/k})/2,0)$ is stable.
There are analogous points on the $s$-axis as  the unstable fixed point
U$_2$:$(0,1-\sqrt{1-4\lambda/k})/2)$ and the stable fixed point
S$_2$:$(0, 1+\sqrt{1-4\lambda/k})/2)$.
On the diagonal line, $r=s$, there are two other unstable fixed points:
U$_3$:$((1-\sqrt{1-8\lambda/k})/4, (1-\sqrt{1-8\lambda/k})/4)$ and 
$U_4$:$((1+\sqrt{1-8\lambda/k})/4, (1+\sqrt{1-8\lambda/k})/4)$. 

If the rate of back reaction $\lambda$ is small, $U_{1,2,3}$ are close to the origin O, 
and U$_4$ and S$_{1,2}$ are close to the diagonal line $r+s=1$.
Then the flow diagram looks similar to Fig.1(c).
However, there is a big difference in the flow diagram Fig.1(c) and (d).
In Fig.1(c) the diagonal line $r+s=1$ is a set of infinitely many fixed points.
On this line, neither $r$ nor $s$ 
varies
 in time.
On the contrary, in Fig.1(d) there are only three stable fixed points;
O and S$_{1,2}$. The trivial fixed point O corresponds to the
structureless or lifeless world, without the chemicals ($R$)-C or ($S$)-C.
Its basin of attraction is a small region enclosed by curves connecting
O and U$_{1,2,3}$.
If the the priomordial environment allows some chemical system 
to reach exterior of this basin, 
chemical complexes with chirality can be produced. 
The nonlinear autocatalysis amplifies the ee,
 and afterwards the back reaction process selects
 only one type of  a chiral component.

\begin{figure}[h]
\begin{center} 
\includegraphics[width=0.8\linewidth]{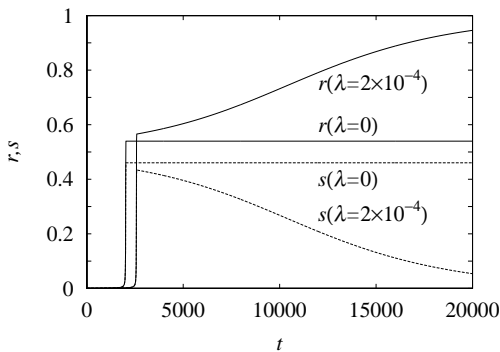}
\end{center} 
\caption{Autocatalytic  evolution of concentrations $r$ (continuous curve)
and $s$ (broken curves) of two enantiomers with ($\lambda=2 \times 10^{-4}$) 
and without ($\lambda=0$) back reaction. 
}
\label{fig2}
\end{figure}

The time evolution of the concentration of each component $r$  and $s$ 
is depicted in Fig.2.
The initial ee is as small as $\Omega_0=0.01$\%
with $r_0=0.00050005 $ and $s_0=0.00049995$.
If there is no back reaction ($\lambda=0$), then in the early stage
reactant concentrations increase as 
$r \approx r_0/(1-r_0t)$ and $s \approx s_0/(1-s_0t)$, and tend to diverge
at times $1/r_0$ and $1/s_0$ respectively.
Before the divergence, the shortage of the reactant A leads to the
saturation of $r$ and $s$, 
as $r \approx r_{\infty}(1-e^{-2kt})$ and $s \approx s_{\infty}(1-e^{-2kt})$.
The final amounts $r_{\infty}$ and $s_{\infty}$ are finite and nonzero.
The ee is amplified to $\Omega_{\infty}=$7.9\% which is less than unity.

If there is a small but a finite back reaction process with $\lambda=0.0002$,
both  $r$ and $s$ increase again and approach close 
to the values corresponding to
the unstable fixed point U$_4$. Then 
the component which has a slight advantage, $r$ in Fig.2, starts to dominate, 
and the other chiral type extinguishes gradually.
The substance ($S$)-C decomposes and supplies the reactant A
which is consummed to increase its enatiomer ($R$)-C. 
The reactant A is, so to say, recycled.
Eventually, the complete homochirality with $\Omega=$100\% is achieved.


We have shown theoretically 
that in a closed chemical system, the nonlinear autocatalysis
amplifies the initial small enantiometric excess.
But eventually, the simple back reaction
promotes the decomposition of less abundant enantiomer to the reactant, which
is recycled to produce the more abundant type.
Through this recycling process, the complete homochirality can be achieved.

The ratio of the forward and the backward reaction rates, 
$k$ and $\lambda$, is related to the equilibrium concentration or the yield. 
Therefore, a system with a yield less than unity might have sufficiently
large back reaction rate to achieve complete homochirality in experimental
time scales.
But if the rate of back reaction is too large and the forward reaction is 
autocatalytic, the system might be trapped into the basin of attraction
of the origin and the reaction can not proceed.
On the contrary, when the rate of back reaction is too small, 
the enatiometric amplification can take place, but the final
selection of chirality should take a long time. 
This might be the reason that the previous experiments by Soai et al.
\cite{sato+01,sato+03}
are well fitted by a model with only the forward autocatalytic reaction.
To observe the effect of back reaction, one might have to wait a long time,
depending on the back-reaction rate.
Otherwise one has to search appropriate chemical systems with proper 
reaction  constants.

If one considers spatial dependence of the chemical reaction,
further complications arise in the process of chirality selection.
Initial productions of enantiomers actually take place randomly 
in space and time.
The nonlinear autocatalysis promotes the chemical reaction
around these initial enantiomers, and
the reaction fronts propagate and expand.
When fronts of different enantiomers meet with each other,
the competition should take place between them.
This competition of enantiomers in space is an interesting future research 
subject.


\end{document}